# Follow-Me Cloud: When Cloud Services Follow Mobile Users

Tarik Taleb, Adlen Ksentini, and Pantelis A. Frangoudis

✦

**Abstract**—The trend towards the cloudification of the 3GPP LTE mobile network architecture and the emergence of federated cloud infrastructures call for alternative service delivery strategies for improved user experience and efficient resource utilization. We propose *Follow-Me Cloud (FMC)*, a design tailored to this environment, but with a broader applicability, which allows mobile users to always be connected via the optimal data anchor and mobility gateways, while cloud-based services *follow* them and are delivered via the optimal service point inside the cloud infrastructure. FMC applies a Markov-Decision-Process-based algorithm for cost-effective, performance-optimized service migration decisions, while two alternative schemes to ensure service continuity and disruption-free operation are proposed, based on either Software Defined Networking technologies or the Locator/Identifier Separation Protocol. Numerical results from our analytic model for FMC, as well as testbed experiments with the two alternative FMC implementations we have developed, demonstrate quantitatively and qualitatively the advantages it can bring about.

## 1 INTRODUCTION

To cope with the explosive growth in mobile data traffic [1], which challenges both their core and radio networks, mobile operators are pushing towards new architectural solutions to decentralize the user plane of their networks. Such approaches involve moving data anchor gateways towards the edge of the network and carefully serving IP traffic via selected points close to their Radio Access Network (RAN) nodes by mobile data offloading techniques [2]. At the same time, computation offloading over heterogeneous wireless network infrastructures also attracts attention, in view of the availability of cloud computing resources [3].

At the service delivery end, the success of cloud computing has led Content/Service Providers to consider deploying more regional Data Centers (DCs). Furthermore, the dependence of content providers (CPs) and Internet Service Providers (ISPs) on one another for efficient content/service delivery and disruption-free network operation in view of dynamic shifts in traffic demands creates CP-ISP cooperation incentives [4] for joint deployment of network-aware content and service delivery infrastructures, and cloud computing technologies are considered for their implementation.

────────────────

• T. Taleb is with Aalto University, Finland.
E-mail: talebtarik@ieee.org
• A. Ksentini and P.A. Frangoudis are with IRISA/Inria Rennes-Bretagne Altantique.
E-mail: adlen.ksentini@irisa.fr, pantelis.frangoudis@inria.fr

Importantly, to more efficiently address the needs of mobile users in terms of geographical coverage and proximity of DCs to them, a new means of cooperative service deployment has emerged in the form of a networked *federated cloud* [5]. This involves allocating virtual resources on a number of DCs dispersed over a specific geographical area, over the infrastructure of potentially heterogeneous federated cloud providers in a transparent manner.

The availability of regional resources and the flexibility of the virtualization technologies upon which federated clouds are built are particularly important to support the modern trend of *cloudifying* the mobile network infrastructure and offering mobile services in an elastic manner, following user demand and presence. In the context of the 3rd Generation Partnership Project (3GPP) Long Term Evolution/Evolved Packet System (LTE/EPS) [6], a decentralized mobile network architecture would include core network gateways such as Packed Data Network Gateways (PGWs) and Serving Gateways (SGWs) operating as *Virtualized Network Function (VNF)* instances on the cloud [7], [8], and not necessarily running on top of specialized dedicated hardware.

At the same time, it is important to ensure that users connected to the mobile core network through a 3GPP base station (eNodeB) or using non-3GPP access, such as Wi-Fi, enjoy acceptable Quality of Experience (QoE) by always guaranteeing optimal end-to-end connectivity for the services offered over the federated cloud, during the entire course of service consumption. Here lies the main challenge we address in this work: While user connectivity to the mobile data anchor gateway is always optimal, this is not necessarily the case for the end-to-end mobile service delivery, since a user on the move may keep receiving the service from a distant (suboptimal) DC after moving to different physical locations.

To answer to this challenge, we introduced the concept of the Follow-Me Cloud (FMC) [9], a design tailored to an interoperating decentralized mobile network/federated cloud architecture. FMC allows not only the content, but also the service itself, to follow a mobile user while moving, ensuring that the latter is always connected to the optimal data anchor and mobility gateways, at the same time accessing a cloudbased service from the optimal DC, in terms either of geographical/network-topological proximity or any other service- or network-level metric, such as load, service delay, etc.

To realize the FMC vision, service continuity and sophisticated schemes for service migration decisions across DCs are critical. In this article, we present a complete framework defining FMC from architectural, algorithmic and implementation perspectives. We explore alternative schemes for ensuring service continuity, which either build on the Locator-Identifier Separation Protocol (LISP) [10], [11] or on using Software-Defined Networking (SDN) technologies. We further propose a Markov Decision Process (MDP)-based algorithm [12] for optimally performing service migration decisions, taking into account user mobility information, and addressing the tradeoff between migration cost and user experience. Our testbed implementation of the proposed architectural alternatives, coupled with an analytic performance evaluation of our system, serve to demonstrate the feasibility and advantages of FMC, and shed light on the practical aspects of its deployment.

The remainder of this article is structured as follows. In Section 2 we provide an overview of related work. We present the FMC concept, entities and high-level functionality in Section 3. Section 4 introduces an analytic model which captures the tradeoff between the benefit and cost due to service migration, and Section 5 presents an MDP scheme building on this model. We describe a LISP-based and an SDN-based implementation of FMC in Section 6, and present analytic and testbed-based performance results in Section 7, before we conclude the article in Section 8.

## 2 RELATED WORK

### 2.1 Service continuity for mobile users

In the mobile networking context that we position our work, a major and well-studied challenge is maintaining service continuity during user and, importantly, service mobility. An approach to this problem is decoupling session and location identifiers. A protocol which makes this separation explicit is LISP (see Section 2.3), and we apply it in this work.

Nordstrom et al. [13], on the other hand, present¨ Serval, a networking stack which includes a new service access layer to cater for user and service mobility, providing identifier/location separation. It makes use of service identifiers, which would however require modifications to legacy applications to support the proposed functionality.

Other research works have considered the use of OpenFlow to hide, through its rules, any changes to IP addresses. OpenFlow-based solutions often face scalability challenges wrt. the number of flows, number of rules, flow setup rate, bandwidth of the control channel, etc. To reduce the number of control packets, DevoFlow [14] moves some of the flow creation work from controllers to switches. Bifulco et al. [15] propose to distribute control plane functions, in order to enhance system scalability, which is an approach that our SDN-based design (see Section 6.1) could follow.

### 2.2 Service migration

In a federated cloud context [5], where geographically distributed DCs are connected into a common resource pool, a cloud management procedure for directing service requests to the optimal DCs, satisfying resource, cost, and quality constraints is necessary. If the respective criteria/constraints are not covered, services may need to be migrated across DCs. Malet and Pietzuch [16] propose a cloud management middleware for migrating part of a user's service (represented by a set of virtual machines) between DC sites in response to DC workload variations and in order to move application components geographically closer to users. Agarwal et al. [17] present Volley, an automatic service placement scheme for geographically distributed DCs based on iterative optimization algorithms, which performs service migrations when detecting that DC capacity or user location change. Alicherry and Lakshman [18] propose a DC selection algorithm for placing a virtual machine (VM), modeling the problem as a sub-graph selection one, while Steiner et al. [19] demonstrate how services can be placed based on information retrieved from an ALTO (Application-Layer Traffic Optimization) server. The above works mainly focus on the VM migration process rather than on VM mobility management.

Other works [20], [21] integrate IP mobility management directly into the hypervisor, which interacts with a VM before and after its migration to update IP addresses in the VM's routing table, or, as in the work of Li et. al [22], invokes Mobile IP functionality each time a VM is created, destroyed or migrated. These solutions perform live VM migration at the expense of potentially long downtimes. Raad et al. [23], on the other hand, achieve sub-second downtimes using a modified version of LISP for rapid traffic redirection. Contrary to our approach (see Section 6.2.2), their scheme also requires modifications to the hypervisor, raising deployment issues.

### 2.3 The Locator-Identifier Separation Protocol (LISP)

With location and identity traditionally coupled, IP mobility becomes a challenging task. To this end, LISP [10] separates them using Routing Locators (RLOCs) and Endpoint Identifiers (EIDs). LISP does not impose any constraints on the EID and RLOC identifier format; IP addresses are typically used. RLOCs are needed to forward packets to/from the Internet, while EIDs are local to an IP subnet. At the data plane level, LISP maps the EID address to an RLOC, and encapsulates the packets into other IP packets before forwarding them through the IP transit. Usually, a LISP site is managed by at least one tunneling LISP router (xTR), having two functionalities: IP packet encapsulation (packet received by a terminal; ingress functionality, or ITR) and decapsulation (packet received by the network; egress functionality, or ETR).

To guarantee EID reachability, the LISP mapping system includes a Map Resolver (MR), a Map Server (MS), and a cache table at each xTR. When a station has a packet to transmit, the EID of the remote station is used in the destination address. Once reaching the ITR (ingress part of xTR), the latter encapsulates the transmitted packet by adding three headers (LISP, UDP, and IP) and fixing the fields "Source Routing Locator" and "Destination Routing Locator" of the LISP header to the source and destination xTR RLOCs, respectively. The mapping between the EID and the corresponding destination xTR RLOC is first looked up in the local cache. If lookup fails, a Map Request message is sent to the Map Resolver, which responds with a Map Reply if the mapping is found. Otherwise, it redirects this request to the Map Server. The MS searches in its local database to find an xTR that would correspond to this EID, and replies with a Map_Reply if it exists. Otherwise, it replies with a Negative Map Reply. Note that the MS receives Map Register messages from ETRs and registers EID-toRLOC mappings in the mapping database.

Compared to mobile IP, LISP avoids triangular routing thanks to decoupling locations and identifiers. A station can move to another location without changing its EID; only the RLOC has to be updated at the MS/MR. Furthermore, with few modifications, LISP can help achieve short VM migration downtimes [23].

### 2.4 Our own prior work

This article extends, generalizes and refines our FollowMe Cloud concept [9], presenting an evolved design targeting generic decentralized mobile network architectures and making heavier use of NFV technologies, bringing the service closer to the end user. We further complement our LISP-based implementation of this scheme [11], which we have updated to match our evolved FMC design, with an SDN-based one. Finally, on the theoretical front, we extend our service migration decision algorithm [12] to also capture 2D mobility scenarios; our algorithm builds on our analytic model presented in [24], included here for completeness.

## 3 THE FOLLOW-ME CLOUD CONCEPT

### 3.1 High-level design

In this section, we present the concept and main functionality of our Follow-Me Cloud (FMC) design for optimized, disruption-free cloud-based services for mobile users. Our high-level architecture is shown in Fig. 1. The two main components of our scheme are the FMC controller (FMCC) and the DC/GW mapping entity. These can either be two independent architectural components, two functional entities collocated with existing nodes, or run as a software on any DC of the underlying cloud.

FMC was designed with the 3GPP LTE/EPS architecture in mind, but is generic and can be applied to other decentralized mobile network access schemes. We assume that DCs are mapped to a set of data anchor routers/gateways. In an LTE context, these routers are PGWs, while, for users roaming across federated Wi-Fi hotspots, such as the Fon network [25], the data anchor router can be the access router of the ISP to which a public Wi-Fi access point is connected. Depending on the mobile access architecture, other options are possible.

The DC-gateway mapping is based on some criterion, such as location or hop-count distance. This mapping may be static or dynamic. In the latter case, topology information can be exchanged between the FMC service provider and the Mobile Network Operator (MNO). Alternatively, an MNO function could be in charge of updating the FMC service provider with such information either in a reactive or a proactive manner. Note, furthermore, that our design includes a single FMCC for managing distributed DC instances, but this does not preclude a decentralized, self-organized implementation for distributed DC coordination.

Taking advantage of virtualization technologies at the data anchor end, our design extends our first version of the FMC architecture [9] with the capability to serve users directly from the data anchor router, bringing services closer to the user end. We distinguish between two types of DCs. At a macroscopic level, there are the core (macro) DCs, which can be considered as data origin servers. At a microscopic level, caches implemented at the data anchor gateways operate as micro-DCs to serve mobile users more efficiently. The focus of the macroDC level is persistent service (VM) storage and service instantiation. The FMC functionality is implemented both at UEs and at the micro-DC level and is responsible for service identification and migration procedures. As shown in Fig. 1, a two-level mapping takes place: MacroDCs are mapped to a group of micro-DCs, each of which is in turn mapped to one or more data anchor routers. Note that these data anchors can be implemented as VNFs hosted in the federated cloud (e.g., collocated with their corresponding micro-DC).

Our design allows for various strategies for selecting which VMs to cache at micro-DCs, as well as for deciding whether a service component will be migrated or replicated at another data anchor following user mobility. However, such strategies are outside the scope of this work; for simplicity and presentation clarity, we assume that a service is deployed at the data anchor router where the user is attached to upon service initiation, and is migrated (i.e., no replication takes place) as a user moves.

From this point on, unless otherwise noted, the term DC will refer to a micro-DC ($\mu$-DC).

## 3.2 Service migration process

With the IP address change which takes place when a UE changes its data anchor router (e.g., PGW reloca-tion in a 3GPP LTE/EPS mobile network architecture), there is a potential need for an FMC service migration. This change can be detected by the serving micro-DC. Whether service migration is worthwhile depends on the service type and requirements (e.g., an ongoing video service with strict QoS requirements may be migrated; a delay-sensitive measurement task for an emergency warning Machine Type Communications service must always be migrated to the optimal DC), content size (e.g., the movie a user is watching is about to finish at the time of PGW relocation; the UE FMC application layer decides not to initiate service migration), and/or user class. The migration decision relies on several potentially conflicting criteria related with user expectations about the service (QoS/QoE, cost) and network/cloud provider policies (load balancing, maximizing DC resource utilization, micro-DC capacity, etc).

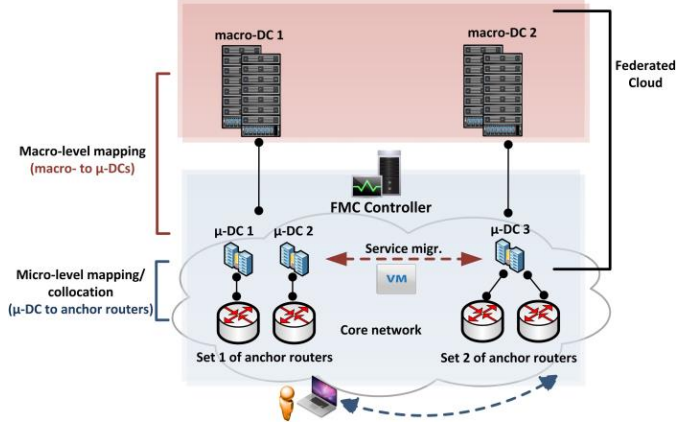

Fig. 1. The FMC high-level architecture in a federated cloud and distributed mobile network environment.

Once the UE or the current micro-DC consider appropriate to migrate the service, the FMC plugin available at the micro-DC may request the FMCC to select the optimal micro-DC to initiate the service migration to. As a service may consist of multiple cooperating components, potentially residing at different locations, the decision has to be made indicating whether the service has to be fully or partially migrated, while considering the service migration cost, e.g., the cost associated with the initiation/replication of a new VM at the target microDC, with the release of resources at the source DC, or with the bandwidth consumption due to traffic being exchanged between the DCs and/or the FMCC. An estimate of these costs shall be compared to the benefits for the (federated) cloud in terms of traffic distribution, but also to those for end users in terms of QoE.

## 4 AN ANALYTIC MODEL FOR FMC

In this section we propose an analytic model to establish the relationship and the relevant tradeoff between FMC service migration cost and benefits in terms of user experience. This model provides insights upon which we base a Markov Decision Process algorithm to derive optimal migration policies (see Section 5).

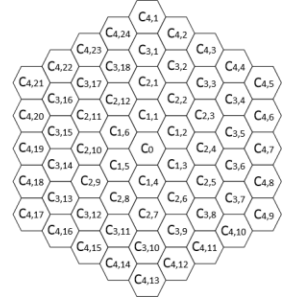

Fig. 2. A typical 3GPP cellular network.

### 4.1 Markov-based system model

We use Markovian models to represent our system, aiming to be able to derive the UE position wrt. the serving DC and predict system evolution. Here, we focus our discussion on a 3GPP LTE mobile network environment. A 3GPP network is typically divided into hexagonal cells (Fig. 2). For the sake of simplicity, we assume that microDCs and data anchor routers (PGWs) are collocated with eNodeBs. In a real implementation, a micro-DC could be mapped to a set of PGWs, which are in turn mapped to a pool of eNodeBs.

We consider a random walk mobility model, where a UE can visit any of the six neighboring cells with probability $p = 1/6$. The residence time of a UE in each cell follows an exponential distribution with mean $1/\mu$. Fig. 2 shows a service area with $k = 5$ rings of cells. Service migration and data anchor gateway relocation are triggered for a UE when its location is $k$ hops away from the serving DC (assumed to be collocated with eNodeBs). Let $X(t)$ denote the distance of the UE to the serving DC (in number of hops) at the time instant $t$. The system $\{X(t), t \geq 0\}$ forms a Continuous-Time Markov Chain (CTMC), with the state space $\{C_{(i,j)} | 0 \leq i \leq (k-1), 1 \leq j \leq 6i\}$.

This chain faces a state space explosion problem, especially if $k$ is high. To address this problem, we reduce the state space by aggregating states that show the same behavior [26], [27], [24], to obtain a new chain $A(t)$ with a lower number of states. In Fig. 2, we see that UEs in ring 0 can move to any neighboring cell with the same probability. UEs in ring 1 come back to the cell which hosts the serving DC with probability $p$, stay in the same ring (same distance from the serving DC) with probability $2p$, and move to ring 2, increasing the distance from the DC, with probability $3p$. Consequently, all states of ring 1 can be aggregated into one state. Regarding ring 2, there are two groups of cells: (i) Cells neighboring three ring-3, two ring-2, and one ring-1 cells, and (ii) cells having two neighbors at each of the 3 rings. Depending on the ring-2 cell the UE is located, e.g., it may either have $3p$ or $2p$ probability to move to ring 3.

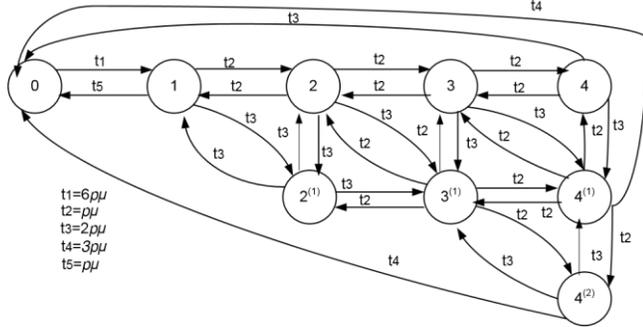

Fig. 3. Markov chain for $k = 5$

Therefore, we obtain two aggregated states: State $C_2$ aggregates states $\{C_{2,1}, C_{2,3}, C_{2,5}, C_{2,7}, C_{2,9}, C_{2,11}\}$ and state $C_2^{(1)}$ aggregates states $\{C_{2,2}, C_{2,4}, C_{2,6}, C_{2,8}, C_{2,10}, C_{2,12}\}$. The same rationale is applied to obtain aggregated states $C_i$ and $C_i^{(m)}$ for any ring $i$, where $1 \leq m \leq \lceil \frac{i-1}{2} \rceil$ is the identifier of an aggregated state within a ring.

As proven by Langar et al. [26], the new aggregated chain $A(t)$, derived from the initial Markov chain $X(t)$, is also Markovian. Fig. 3 shows the transition diagram of the aggregated Markov chain when the service migration is triggered when the UE is $k = 5$ hops away from the serving DC. Based on this figure, we can derive the steady state probabilities of the aggregated states. For simplicity, each aggregated state of ring $i$ in Fig. 3 is labeled using the ring number and the superscript used to identify different aggregate states of the same ring, if necessary. The balance equations (Eq. (1)-(6)) to solve the system follow[1]:

$$\begin{cases} \pi_0 = \frac{1}{6}\pi_1 + \frac{1}{2}\pi_{k-1} + \frac{1}{3}\sum_{j=1}^{\lceil \frac{k-2}{2} \rceil} \pi_{k-1}^{(j)} \\ \pi_1 = \pi_0 + \frac{1}{3}\pi_1 + \frac{1}{6}\pi_2 + \frac{1}{3}\pi_2^{(1)} \\ \pi_2 = \frac{1}{6}\pi_1 + \frac{1}{6}\pi_3 + \frac{1}{3}\pi_2^{(1)} + \frac{1}{6}\pi_3^{(1)} \\ \pi_{k-1} = \frac{1}{6}\pi_{k-2} + \frac{1}{6}\pi_{k-1}^{(1)} \\ \pi_i = \frac{1}{6}\pi_{i-1} + \frac{1}{6}\pi_{i+1} + \frac{1}{6}\pi_{i-1}^{(1)} + \frac{1}{6}\pi_{i+1}^{(1)}, \forall 3 \leq i \leq k-2 \end{cases} \quad (1)$$

$$\begin{cases} \pi_2^{(1)} = \frac{2}{3}\pi_1 + \frac{1}{3}\pi_2 + \frac{1}{6}\pi_3 \\ \pi_3^{(1)} = \frac{1}{3}\pi_2 + \frac{1}{3}\pi_3 + \frac{1}{3}\pi_2^{(1)} + \frac{1}{6}\pi_3^{(1)} + \frac{1}{6}\pi_4^{(1)} + \frac{1}{3}\pi_4^{(2)} \\ \pi_4^{(1)} = \frac{1}{3}\pi_3 + \frac{1}{3}\pi_4 + \frac{1}{6}\pi_3^{(1)} + \frac{1}{6}\pi_5^{(1)} + \frac{1}{3}\pi_4^{(2)} + \frac{1}{6}\pi_5^{(2)} \\ \pi_{k-1}^{(1)} = \frac{1}{3}\pi_{k-2} + \frac{1}{3}\pi_{k-1} + \frac{1}{6}\pi_{k-2}^{(1)} + \frac{1}{6}\pi_{k-1}^{(2)} \\ \pi_i^{(1)} = \frac{1}{3}\pi_{i-1} + \frac{1}{3}\pi_i + \frac{1}{6}\pi_{i-1}^{(1)} + \frac{1}{6}\pi_{i+1}^{(1)} + \frac{1}{6}\pi_i^{(2)} + \frac{1}{6}\pi_{i+1}^{(2)} \\ \qquad \forall 5 \leq i \leq k-2, \end{cases} \quad (2)$$

$$\begin{cases} \pi_i^{(j)} = \frac{1}{6}\pi_i^{(j-1)} + \frac{b_1}{6}\pi_i^{(j+1)} + \frac{1}{6}\pi_{i-1}^{(j-1)} + \frac{1}{6}\pi_{i-1}^{(j)} \\ \quad + \frac{b_2}{6}\pi_{i+1}^{(j)} + \frac{b_2}{6}\pi_{i+1}^{(j+1)}, \quad 2 \leq j \leq \lceil \frac{i-1}{2} \rceil - 1 \\ \qquad \forall 6 < i < k-1 \text{ and} \end{cases} \quad (3)$$

[1]. For details on the state aggregation algorithm and more insight on the derivation of the balance equations for the reduced CTMC, the reader is referred to the work of Langar et al. [26].

where

$$b_1 = \begin{cases} 1 & \text{if } i \text{ is odd} \\ 1 & \text{if } i \text{ is even and } 2 \leq j \leq \lceil \frac{i-1}{2} \rceil - 2 \\ 2 & \text{if } i \text{ is even and } j = \lceil \frac{i-1}{2} \rceil - 1 \end{cases}$$

and

$$b_2 = \begin{cases} 0 & \text{if } 6 \leq i \leq k-2 \\ 1 & \text{if } i = k-1 \end{cases}$$

$$\begin{cases} \pi_{2l}^{(l)} = \frac{1}{6}\pi_{2l}^{(l-1)} + \frac{1}{6}\pi_{2l-1}^{(l-1)} + \frac{c_1}{6}\pi_{2l+1}^{(l)}, \quad \forall 2 \leq l \leq \lceil \frac{k-1}{2} \rceil \end{cases} \quad (4)$$

where

$$c_1 = \begin{cases} 0 & l = \frac{k-1}{2} \\ 1 & \text{otherwise} \end{cases}$$

$$\begin{cases} \pi_{2l+1}^{(l)} = \frac{1}{6}\pi_{2l+1}^{(l-1)} + \frac{1}{6}\pi_{2l+1}^{(l)} + \frac{1}{6}\pi_{2l}^{(l-1)} + \frac{1}{6}\pi_{2l}^{(l)} \\ \quad + \frac{c_2}{6}\pi_{2l+2}^{(l)} + \frac{c_2}{6}\pi_{2l+2}^{(l+1)}, \quad \forall 2 \leq l \leq \frac{k-2}{2} \end{cases} \quad (5)$$

where

$$c_2 = \begin{cases} 0 & \text{if } l = \frac{k-2}{2} \\ 1 & \text{otherwise} \end{cases}$$

$$\sum_{i=0}^{k-1} \pi_i + \sum_{i=2}^{k-1} \sum_{m=1}^{\lceil \frac{i-1}{2} \rceil} \pi_i^{(m)} = 1 \quad (6)$$

### 4.2 Average UE-DC distance and the probability to be connected to the optimal DC

Let $E[Dist]$ denote the average distance of a UE from the serving DC. $E[Dist]$ depends on the value of $k$, and the distance (number of hops) of the UE from the data anchor router collocated with the serving DC. Recall that a UE remains connected to this anchor and all data are consequently routed through the latter until service migration is triggered. Therefore, the average distance is expressed as:

$$E[Dist] = \sum_{i=1}^{k-1} i\pi_i + \sum_{i=2}^{k-1} \sum_{j=1}^{\left\lceil \frac{k-2}{2} \right\rceil} i\pi_i^{(j)}. \quad (7)$$

The probability that the UE is connected to the optimal DC during the system's lifetime is $\pi_0$.

### 4.3 Average end-to-end delay from the serving DC

We define the end-to-end (e2e) delay as the delay for a UE to receive data packets from the serving DC. Similar to $E[Dist]$, the e2e delay depends on the UE distance (number of hops) to the data anchor router connecting to the DC. The average e2e delay is denoted by $E[D]$ and is given by:

$$E[D] = \sum_{i=1}^{k-1} D_i \pi_i + \sum_{i=2}^{k-1} \sum_{j=1}^{\left\lceil \frac{k-2}{2} \right\rceil} D_i \pi_i^{(j)}, \quad (8)$$

where $D_i$ is the e2e delay when the UE is at distance $i$ (cells belonging to ring $i$).

### 4.4 Service migration cost

$MC$ denotes the cost of migrating part (i.e., some of the components/sessions composing it) or all the service from a DC to the optimal one. It depends on the size of the objects to be migrated, as well as the amount of signaling messages exchanged among the FMCC, the UE and the DCs. In FMC, there are three signaling messages to trigger service migration. The cost for a service migration thus follows:

$$Cost = Objects_{size} + 3SIG_{size}, \quad (9)$$

where $SIG_{size}$ is the signaling message size. Hence, $MC$ can be derived as follows:

$$MC = \left[ 3p\pi_{k-1} + 2p \left( \sum_{j=1}^{\left\lceil \frac{k-2}{2} \right\rceil} \pi_{k-1}^{(j)} \right) \right] \times Cost. \quad (10)$$

### 4.5 Service migration duration

The service migration duration is the time required to transfer part or all of the service from the current DC to the optimal one. It mainly depends on: (i) the size of the objects to transfer; (ii) the RTT of the TCP connection between the two DCs; and (iii) the time needed to convert a VM to the appropriate format, if the two DCs are not using the same hypervisor. It also represents the time when the service cannot be used, in other words, service disruption time (denoted as $SDT$). Assuming that the data transfer is based on an FTP-like application, we use the empirical TCP latency model of Sikdar et al. [28], and the $SDT$ value can be computed as follows:

$$SDT = [\log_{1.57} N + f(p_{loss}, RTT)N + 4p_{loss} \log_{1.57} N$$
$$+ 20 p_{loss} + \frac{(10 + 3RTT)}{4(1 - p_{loss})W_{max}\sqrt{W_{max}}} N] RTT$$
$$+ T_{VM\_conversion} \quad (11)$$

where $p_{loss}$ denotes the packet loss rate, $N$ is the number of packets to transfer, $W_{max}$ is the maximum size of the congestion window, $T_{VM\ conversion}$ is the time required to convert a VM and

$$f(p_{loss}, RTT) = \frac{32(2p_{loss}^2 + 4p_{loss}^2 + 16p_{loss}^3)}{(1+RTT)^3} N + \frac{(1+p_{loss})}{RTT10^3}\ 2.$$

Note that $N = \left\lceil \frac{Service_{size}}{MSS} \right\rceil$, where MSS is the maximum segment size used by the TCP connection.

## 5 A MDP-BASED SCHEME FOR SERVICE MIGRATION

We model the service migration decision as a Markov Decision Process (MDP), capturing the tradeoff between reducing cost and maintaining satisfactory user experience. This model decides whether a service consumed by a user at distance $d$ from the current DC should be migrated to an optimal DC, a decision process carried out by the FMCC. To formulate the service migration decision policy, we define a Continuous Time Markov Decision Process (CTMDP) that associates to each state an action, corresponding transition probabilities, and rewards.

Let $s_t$ be the process describing the evolution of the system state and $S$ denote the state space. We assume the cellular network topology model of Fig. 2. Each cell belongs to a Tracking Area (TA) and each TA belongs to a Service Area (SA), which is served by one anchor gateway (PGW or access router). Fig. 4 shows the CTMDP for the case of a one-dimensional (1D) mobility model: A mobile user has only two possible destinations, i.e., a new SA with probability $p$, or moving back to a visited SA with probability $1 - p$. Higher values of $p$ indicate that a user is moving far from the current DC. Fig. 5 illustrates the case for the 2D mobility model described in the previous section. The vector $A = (a_1, a_2)$ describes the actions available to the FMCC at each epoch (i.e., when a UE performs handoff and enters into another SA). Action $a_2$ is used if the service is migrated to an optimal DC, while action $a_1$ is used if the UE is still served by the same DC. Depending on the current state, the set of available actions differs. For the sake of simplicity, we demonstrate the use of MDP to solve the service migration problem for 1D mobility. The same reasoning can be applied to the case for the 2D model shown in Fig. 5. Note that, albeit its simplicity, the 1D model is appropriate for vehicular networking environments where users move along predefined trajectories, such as highways or railway tracks.

In the 1D mobility model, the residence time of a user in each SA follows an exponential distribution with mean $1/\mu$. Hence, the state space $S$ is defined as $S = \{0,1,...,thr\}$, where *thr* represents the maximum distance (in terms of visited SAs) from where the service must be migrated to the optimal DC.

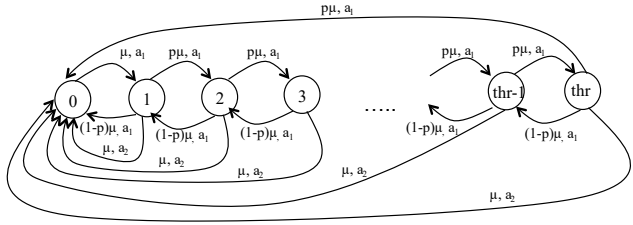

Fig. 4. CTMDP of the service migration procedure: 1D mobility model.

The FMCC observes the current state $s$ of the network and associates a set of possible actions $A_s$ to it, taken upon arrival to it from the previous state. For a given action $a$, an instantaneous reward $r(s,s^0,a)$ is associated to a transition from state $s$ to another state $s^0$. The corresponding formal representation of the CTMDP is as follows:

$$(S, (A_s, s \in S), q(s^0|s,a), r(s,s^0,a)).$$

For particular states, the set of possible actions $A$ reduces to a subset $A_s$. A policy $P$ associates an action $a(s|P)$ to a state $s$. Let $Q$ be the transition matrix, with

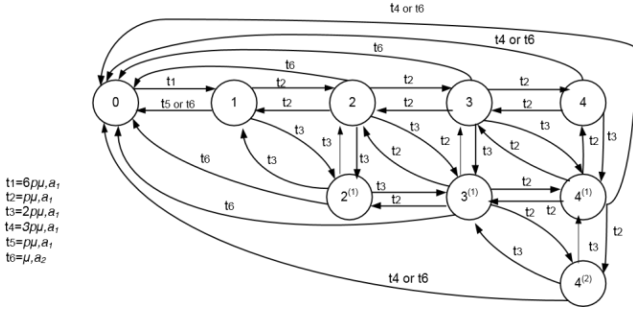

Fig. 5. CTMDP of the service migration procedure: 2D mobility model.

$q(s^0|s,a)$ denoting the transition rate between states $s$ and $s^0$ in $S$ due to action $a$, which, in the FMC scenario, represents a UE moving from one SA to another SA. By construction, we define a policy as a function of the actual state. The decision to migrate a service or not is taken by observing only the actual state. Since this process is Markovian (the sojourn time in a SA follows an exponential distribution), the controlled process is also Markovian. To resolve the MDP, we use an equivalent Discrete Time Markov Decision Process (DTMDP) for the defined CTMDP, with a finite state space $S$. We argue that the state space is finite, since, after a certain distance (*thr*) from the current DC, the service is automatically migrated to the optimal DC. For each $s \in S$, $A_s$ represents the finite set of allowed actions in that state. This DTMDP can be derived by uniformization and discretization of the initial process as follows: When all transition rates in matrix $Q$ are bounded, the sojourn times in all states are exponential with bounded parameters $tr(s|s,a)$. Therefore, a $sup_{(s \in S, a \in A_s)} tr(s|s,a)$ exists and there is a constant value $c$ such that $sup_{(s \in S, a \in A_s)}[1 - p(s,a)] tr(s|s,a) \le c < \infty$,

where $p(s|s,a)$ denotes the probabilities of staying in the same state after the next event. We can now define an equivalent uniformized process with state-independent exponential sojourn times with parameter $c$, and transition probabilities:

$$p(s'|s,a) = \begin{cases} 1 - \frac{([1-p(s|s)]tr(s|s,a))}{c} & s = s' \\ \frac{p(s'|s)tr(s'|s,a)}{c} & s \ne s' \end{cases}.$$

By setting $c = \mu$, the transition probabilities of the DTMDP procedure are defined as follows:

$$p(j|s,a) = \begin{cases} 1 & j = 0, s \ne 0, s \ne thr, a = a_1 \text{ or } j = 1, s = 0, a = a_2 \\ p & j = s+1, s \ne 0, s \ne thr, a = a_1 \\ & \text{or } j = 0, s = thr, a = a_1 \\ & j = s-1, s \ne 0, a = a_1 \\ 10 - p & \text{Otherwise} \end{cases}.$$

It is important to note that when the system is in state $s = thr$, the only available action is $a_1$. If the UE moves to another SA, where the distance exceeds the threshold *thr*, service migration is automatically triggered.

In the remainder of this section, we use the DTMDP version. For $t \in N$, let $s_t$, $a_t$ and $r_t$ denote the state, action and reward at time $t$ of the DTMDP procedure, respectively. Let $P^a_{(s,s')} = p[s_{(t+1)} = s_0 | s_t = s, s_{(t+1)} = s_0, a_t = a]$ denote the transition probabilities and $R^a_{(s,s')} = E[r_{(t+1)} | s_t = s, s_{(t+1)} = s', a_t = a]$ denote the expected reward associated with the transitions. A policy $\pi$ is a mapping between a state and an action, and can be denoted as $a_t = \pi(s_t)$, where $t \in N$. Accordingly, a policy $\pi = (\theta_1, \theta_2, \theta_3, ..., \theta_N)$ is a sequence of decision rules to be used at all decision epochs. We restrict ourselves only to deterministic policies, as they are simple to implement [29]. When a UE hands off a particular SA to another SA, the FMCC has to decide either to migrate the service using action $a_2$ or not to migrate it using action $a_1$. For each transition, a reward is obtained. This reward is a function of the cost of migrating a service (zero in case of no migration) and the quality obtained from the new state. The cost of migrating a service is defined as follows:

$$g(a) = \begin{cases} 0 & a = a_1 \\ C_m & a = a_2 \end{cases}$$

where $C_m$ denotes the cost of migrating all the service or a part of it. Therefore the reward function is given by $r(s,s^0,a) = Q(s^0) - g(a)$, where $Q(s)$ quantifies the quality perceived by a user connected to the source DC when at state $s$. Note that quality is inversely proportional to the hop distance from the source DC, and $Q(0)$ is the maximum quality that a UE can enjoy, when connected to the optimal DC. In general, the quality function can be expressed in the form $Q(s) = Q(0) - K$, where $K$ denotes a predetermined factor. Given a discount factor $\gamma \in [0,1]$ and an initial state $s$, we define the total discount reward policy $\pi = (\theta_1, \theta_2, \theta_3, ..., \theta_N)$ as

$$v_\gamma^\pi = \lim_{N \to \infty} E_\gamma^\pi \{\sum_{t=1}^N \gamma^{t-1} r_t\} = E_\gamma^\pi \{\sum_{t=1}^\infty \gamma^{t-1} r_t\}.$$

Given the uniformization of the CTMDP, $r(s,s^0,a)$ explicitly depends on the transitions between states. The new reward function $r^0(s,s^0,a)$ is obtained as follows [29]:

$$r'(s,s',a) = r(s,s',a) \frac{\alpha + \beta(s,s',a)}{\alpha + c},$$

where $\beta(s,s^0,a)$ is the transition rate between states $s$ and $s^0$ when using action $a$, and $\alpha$ is a predetermined parameter. With the new formulation of the reward function and the uniformization of CTMDP, we can use the discounted models as in discrete models to resolve the system [29]. Let $v(s) = max_{\pi \in \Pi} v^\pi(s)$ denote the maximum discounted total reward, given the initial state $s$. From [29], the optimality equations are given by $v(s) = max_{\pi \in \Pi} \{r^0(s,s^0,a) + \sum_{s^0 \in S} \gamma P[s^0|s,a] v(s^0)\}$.

The solutions of the optimality equations correspond to

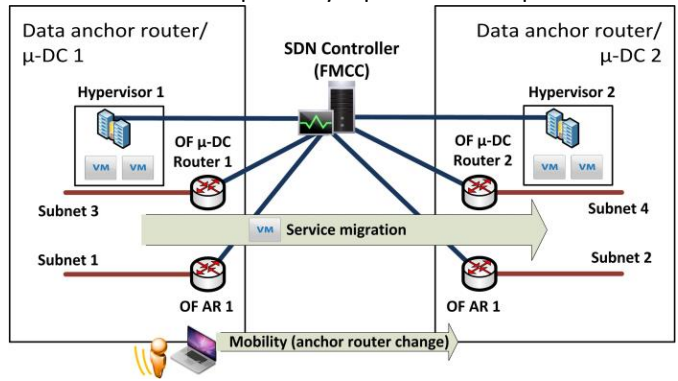

## 6 IMPLEMENTATION ALTERNATIVES

To show the breadth of available options to realize the FMC architecture, we explore alternative enabling technologies for its implementation. We have identified two candidate approaches, one based on SDN and one on the LISP protocol. Note however that, as we have shown [30], the FMC design can be integrated also with the Mobile IPv6 protocol.

### 6.1 An OpenFlow-based FMC architecture

#### 6.1.1 Design

Our SDN-based scheme is built on the NOX OpenFlow Controller [31] and is shown in Fig. 6. Mobile users access cloud-based services over a (mobile/wireless) client network and their traffic is forwarded by OpenFlowcapable access routers (OF AR x). Traffic from/to each federated cloud location (data center), inside which VM instances interconnected via virtual switches are managed by local hypervisors, is routed through OpenFlow micro-Data Center routers (OF $\mu$-DC x). At the core of this architecture is our NOX-based FMCC, with which the components of our architecture communicate. The controller is assumed to be aware of i) the virtual switch instances and their data path identifiers on the physical OpenFlow switch, ii) the VM identifiers (namely the IP and MAC addresses), iii) the location and IP addresses of each default gateway in the topology, iv) the OpenFlow switch port identifiers at which the DC, router and client networks are connected, v) the IP address ranges managed by each DHCP server both for client and

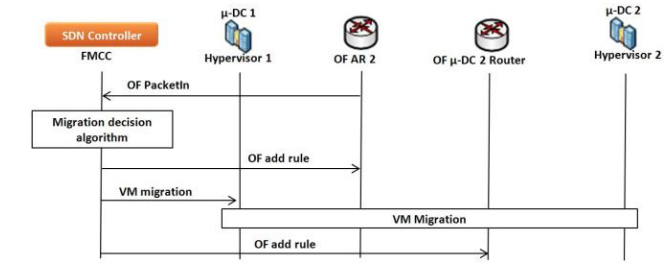

Fig. 7. Service migration procedure in an SDN-based FMC implementation.

Fig. 6. OpenFlow-based Follow-Me Cloud architecture.

the maximum expected discounted total reward $v(s)$ and the optimal policy $\pi^*(s)$. It is worth mentioning that the optimal policy $\pi^*(s)$ indicates the decision as to which network and which DC the UE is to be attached and to be connected, respectively, given the state $s$. There are several algorithms that can be used to solve the optimization problem given by the above optimality equations. Value iteration and policy iteration

DC networks, and vi) the locations of distributed data centers that can either be part of the operator network or could be autonomous domains. The basic functionalities of the SDN-based FMCC follow.

**Location Management:** For correctly installing forwarding rules into the OpenFlow switch, each client and VM are linked to home locations, based on their IP address allocation and gateway settings. The current location of a client in the client network is also maintained. If any traffic from a particular client or VM appears on a different network than the one corresponding to its home location, the FMCC updates the status of that entity to indicate that it is in a Visited Network/Location.

The location management process is also responsible for selecting the appropriate micro-DC location for the VM serving a client, utilizing information about the geographic location of DCs and characteristics of the client-VM/DC network paths, such as average delay and network load or congestion.

**Mobility detection and service migration:** The actual VM migration is carried out by the cloud infrastructure software. (See Section 6.1.2 for some details on our proofof-concept implementation.) The sequence of messages and events is shown in Fig. 7.

When a user changes its point of attachment, the OpenFlow Access Router on the visited network sends a PacketIn OpenFlow message to the controller, which is thus notified of the mobility event. This process can be initiated when a user performs the DHCP message exchange with the visited network. Then, the FMCC executes the service migration decision algorithm (see Section 5), and, if migration is necessary, it installs the appropriate OF rule at the OF AR of the visited network and launches the VM migration, also adding a rule to the OpenFlow router serving the micro-DC which hosts the migrated VM, so that when traffic for the latter reaches the OpenFlow switch of a visited network, it can be matched in the flow table of that switch.

**Session management:** A key functional requirement is for the controller to preserve all ongoing user sessions while a VM is migrated. This implies that no configuration change (e.g., IP address and gateway configurations) is allowed on the VM. Furthermore, the (private) IP address ranges managed by DCs can be overlapping, and the first-hop setting may not be consistent across subnet boundaries. We address this by setting up a tunnel within the visited network segment. The tunnel operates by rewriting the IP address field within the packet IP header for each outgoing packet from the VM to the external network. The original IP header is restored in the packet when the last hop in the visited network segment is reached. The same technique is applied for all the incoming traffic to the VM. This is achieved by modifying the set of OpenFlow rules installed in the visited network.

### 6.1.2 Experimental setup

We have developed an experimental testbed implementing the SDN-based architecture of Fig. 6. Our setup includes two DCs, each one implemented as a single VMWare ESXi hypervisor. Each ESXi host is equipped with two 1 Gbps Ethernet cards for forwarding management and OpenFlow traffic over the network. A virtual network topology is defined inside the ESXi host by two vSwitches (soft switches), where each physical NIC is connected with each soft switch instance. The ESXi host manages Windows XP VMs and each VM is configured with two virtual NICs (vNIC) connected with the virtual network through the soft switches. One vNIC carries management traffic (vmKernel) and the other carries OpenFlow traffic. The storage space is shared between the two DCs and is accessed by the standard iSCSI protocol. The DCs are remotely managed by the VMWare vCenter software.

There are two separate WLANs for client connectivity and two Linux-based hosts act as first-hop routers for client traffic, assigning client addresses using DHCP and running Linux traffic control (tc) for controlling path characteristics (e.g., delay and available bandwidth, and thus congestion) between the two network segments. In this simplified setup, each DC host plays the role of a micro-DC mapped to a data anchor router (in our case the WLAN first-hop router), and is placed in each of the two client networks (and served by the respective router).

The NOX-based FMCC, as specified in Section 6.1, the Linux-based routers, ESXi hosts and the VMWare vCenter host are all connected to ports of an NEC IP-8800 OpenFlow switch. From the physical OpenFlow switch perspective, four virtual switches (VLAN) are used for separately carrying the traffic of the two data centers and the client network. The FMCC manages the forwarding behavior on the four VLAN's and also monitors the path characteristics between a DC and the client network for resource management optimizations. For triggering live VM migration across DCs, the VMotion cloud infrastructure technology and proprietary API from VMWare are used. VMotion traffic is mapped to the management network, while all active communication between the VM and remote users is managed by the OpenFlow network.

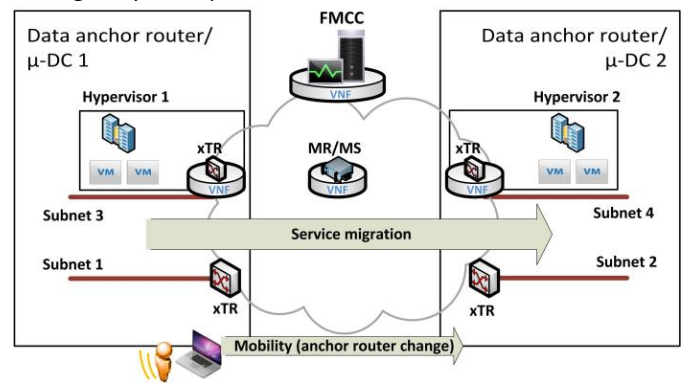

Fig. 8. A LISP-based Follow-Me Cloud architecture.

## 6.2 A LISP-based FMC architecture

### 6.2.1 Design

Our second architectural alternative is based on the LISP protocol. Each micro-DC is connected to the Internet through an xTR router. The client (mobile) network domain contains IP subnets interconnected through xTR routers as well, and the architecture includes a LISP MR/MS element. All LISP entities (MR/MS and xTRs) are implemented as VNFs and are deployed on VMs in the cloud. Note that, in practice, the xTR routers of the mobile transport network domain could either be VNFs or be built directly on top of the data anchor router hardware. This FMC architecture also includes an FMCC in the form of a VNF.

Fig. 8 shows the envisioned LISP-based FMC architecture and scenario. A mobile user accessing a service hosted at micro-DC1, and initially connected to Subnet 1, moves to Subnet 2. The xTR router of Subnet 2 notifies the MR/MS about this mobility event. The MR/MS updates its cache and informs the FMCC about the new location of the user. The FMCC executes the service migration decision algorithm (see Section 5) to decide whether to migrate the user's service (that is, the VM(s) hosting the service) to the micro-DC corresponding to the new location of the user. If the decision is positive, the FMCC requests Hypervisor 1 to launch the migration procedure. Since the VM is migrated to Hypervisor 2, the xTR router of Subnet 4 is informed of the migration event, and the latter further informs the MR and the xTR router of Subnet 3 (as well as other xTR routers involved in communication with the VM) accordingly. Finally, the MR/MS resolver updates its cache and notifies the FMCC about this change.

### 6.2.2 Support for service continuity

To ensure service continuity, a LISP-assisted live service migration mechanism should be capable of (i) maintaining the VM Endpoint identifier (EID) when migrating it from its current DC to the target one, (ii) updating the Routing Locator (RLOC) of the target xTR router to include the VM's EID, (iii) informing the MR server and all xTR routers involved in the communication with the migrated VM to update the RLOC of the migrated VM, and (iv) informing the old xTR router to erase the VM EID from its cache.

In this work, we assume that a VM's EID is the first IP address it obtains, and that its RLOC is the IP address of the corresponding xTR router. Furthermore, we consider that a VM's EID is registered at the initial xTR router with a large IP subnet prefix (in our case, /24). The EID is mapped to the RLOC of the source xTR router at the MR, as well as in the caches of xTR routers communicating with the VM.

When the VM is migrated to the target hypervisor, the EID of the migrated VM should be maintained and the destination xTR router has to be informed about the migration event. Different approaches exist to achieve this. In one approach [32], the xTR router does not become aware of the new VM until the latter initiates communication, when the xTR detects that the source IP address (migrated VM's EID) is not belonging to its IP subnet.

Although this solution does not require any signaling messages, it can break the current VM's connections and thus does not ensure service continuity; if the VM has no packet to transmit, the current xTR router communicating with the VM may continue using its old Routing Locator (RLOC). An alternative to this approach was proposed by Raad et al. [23], where LISP is used in the control plane to inform the source and target xTR routers about a successful VM migration. A new Change Priority (CP) LISP message is introduced, which allows (i) the target hypervisor to inform the target xTR router about the migration of a new VM to the target DC (including the VM's EID), and (ii) to update the cache (RLOC-EID mapping) of other xTR routers. However, this solution requires modifying the hypervisor, making it hard to implement and deploy, as the hypervisor software is independent from the operator (as well as the LISP domain).

In this article, we propose another approach (Fig. 9), where the FMCC informs both xTR routers (handling the involved DC domains) about the change in the VM's RLOC. Since the FMC controller is integrated in the LISP domain (it already communicates with MR/MS to track user locations), it could easily be made aware of the xTR router handling a DC domain. This could be obtained by sending a message to MR/MS to know the xTR router handling the DC's IP domain.

Once the xTR router becomes aware of the migration of a new VM to its local DC, it notifies the MR/MS to update its RLOC by including the migrated VM's EID. The EID of the migrated VM is in the form of the initial IP address, but with a /32 prefix. Therefore, the RLOC of the target xTR router is mapped to both its subnet prefix and the VM's EID prefix (/32). Furthermore, the former xTR router erases the old EID-to-RLOC entry from its cache. To speed up traffic redirection, the source xTR router uses a new LISP message (as in [23]) to inform the other xTR router which was communicating with the concerned VM so that the latter accordingly updates the VM's RLOC. Note that the xTR router should maintain a list of xTR routers involved with each active connection.

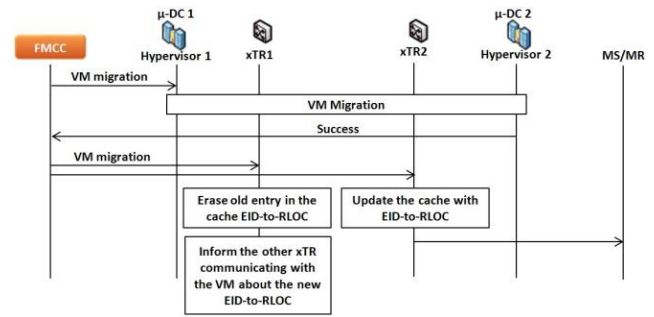

Fig. 9. Service migration procedure in a LISP-based FMC implementation.

### 6.2.3 Experimental setup

The VNFs corresponding to the entities of our LISPbased FMC scheme are implemented using the Click modular router framework [33] and run as ClickOS [34] VMs deployed on a Xen hypervisor. DCs are emulated using kvm [35] to take advantage of the VM migration options that it offers, as kvm can migrate only the RAM content between the involved DCs, while the hypervisors have a shared storage via NFS. We emulate a mobile user moving between two IP domains; whenever the user enters a new domain, service (VM) migration is triggered. Each VM is Ubuntu 8.04, with 1 GB of RAM and a one-core processor. DCs (kvm managers) run Ubuntu 14.04 with 8 GB of RAM. We emulated varying latencies in the paths between DCs and between the FMCC and xTR routers using Linux tc.

## 7 PERFORMANCE EVALUATION

In this section we present a quantitative performance evaluation of the proposed scheme. We begin with numerical results from our analytic model, followed by measurements carried out on our testbed implementation of the SDN- and LISP-based FMC solutions.

### 7.1 Model-based performance results

We present numerical results obtained by resolving the Markov model defined in Section 4.1. We evaluate the performance of FMC in terms of the probability that the UE is connected to the optimal DC, the average distance of the UE from the optimal DC, the UE connection latency, the service migration cost and the service disruption time during service migration.

We assume a reliable connection between DCs (zero packet loss), the total size of the service to migrate (i.e., the respective VM) is set to 1 Gb, and all DCs are assumed to use the same hypervisor; thus, there is no VM conversion cost when migrating a service. Note that the case for $k = 7$ corresponds to a situation where the FMC concept is not used, as $k$ is unrealistically high; in practice, the service area is typically limited to a modest value for k.

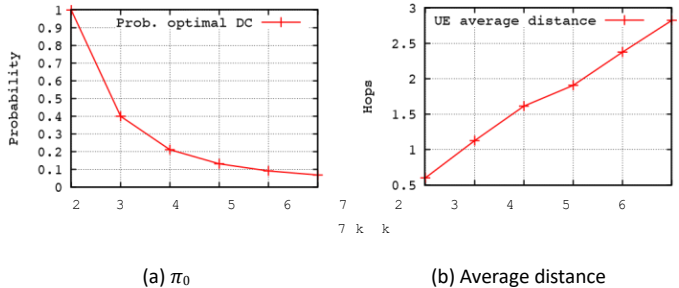

(a) $\pi_0$  (b) Average distance

Fig. 10. Probability to be connected to the optimal DC and the average distance from it.

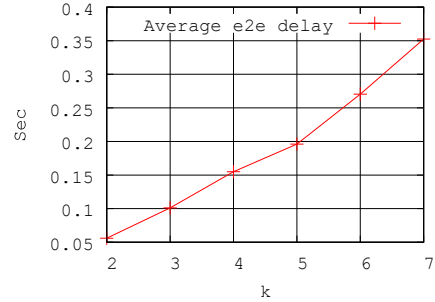

Fig. 11. Average latency

Fig. 10(a) and 10(b) show the probability of a UE to be connected to an optimal DC, and the average distance from it for different values of $k$, respectively. We notice that this probability is a decreasing function of $k$: High probability is obtained when the service migration is triggered after each UE handover, ensuring that the UE is always connected to the optimal DC, while the opposite effect appears when delaying service migration to longer distances. On the other hand, the average distance is an increasing function of $k$. Indeed, if service migration is delayed, the UE is likely connected from a distance higher than one hop. This average distance exceeds two hops when $k$ is higher than 6.

In Fig. 11, we plot the average latency of the UE connection for different values of $k$. Note that latency is given by $Lat_i = 0.02i^2$ (s), where $i$ is the hop distance from the serving DC. Intuitively, the average latency increases with $k$. If we compare the cases of using $k = 2$ and $k = 7$, we see a significant difference of about 200 ms in delays.

Fig. 12 shows the service migration cost for different values of $k$. We present results for migrating all, 50%, and 10% of the service. This cost is a decreasing function of $k$, and it is high when service relocation is launched at each UE handover. Furthermore, the highest cost is reached when migrating all the service, as it critically depends on the object size to migrate.

Fig. 13 depicts the service disruption time for different values of $k$. We assume that RTT is proportional to the square of the hop distance $i$ between two DCs, given by $RTT_i = 0.01i^2$ (s). Similar to Fig. 12, we considered three service migration cases: migrating all, 50%, or 10% of the service. We clearly observe that the SDT value is an increasing function of $k$. This is attributable to the fact that high values of $k$ mean longer distances between the concerned DCs, thus an increased value of RTT. In addition, we notice that the SDT value is the highest when migrating all the service, mainly due to the larger size of the objects to transfer. Notably, this difference is not important for low values of $k$. Since SDT is highly dependent on the RTT, accelerating data transfers using tools like FDT [36] is mandatory when the RTT is high.

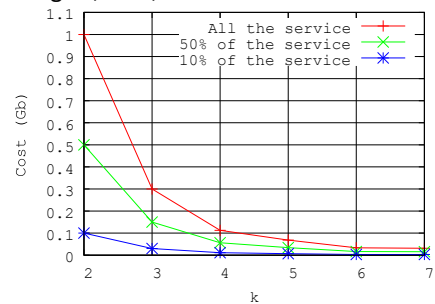

Fig. 12. Cost of service migration.

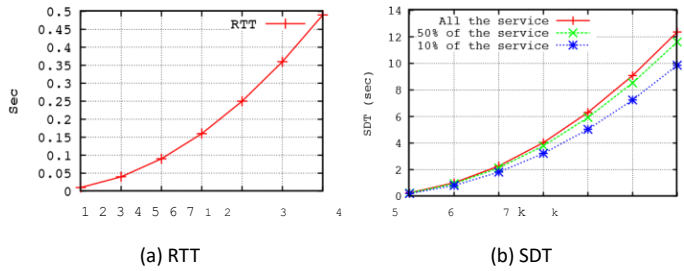

(a) RTT   (b) SDT

Fig. 13. Service disruption time – RTT proportional to the square of the distance.

### 7.2 Service migration policies

Given the tradeoff between migration cost and user experience improvement identified in Section 7.1, we demonstrate the construction of service migration policies obtained using a Matlab implementation [37] of the Value Iteration algorithm [29]. We set $thr = 10$, i.e., a service migration is automatically triggered if the UE is at a distance higher than 10 (in terms of the number of visited SAs) from the source DC. The factor $K$ of the quality function is arbitrarily set to 1. We introduce a new metric $\tau$ representing the ratio between the cost and the maximum quality $Q(0)$. Two scenarios are studied:

- $\tau = 0.1$, which represents a low cost compared to the quality obtained if a service migration is launched. This could be the case when only a part of a service is migrated.
- $\tau = 0.5$, which indicates that the cost is not negligible compared to the quality obtained if a service migration is launched.

Fig. 14(a) and 14(b) illustrate the optimal policy constructions for the aforementioned scenarios. The inter-

| p\d | 1 | 2 | 3 | 4 | 5 | 6 |
|---|---|---|---|---|---|---|
| 0.1 | C | C | C | C | C | C |
| 0.2 | C | C | C | C | C | C |
| 0.3 | C | C | C | C | C | C |
| 0.4 | C | C | C | C | C | C |
| 0.5 | C | C | C | C | C | M |
| 0.6 | C | C | C | C | C | M |
| 0.7 | C | C | C | C | M | M |
| 0.8 | C | C | C | C | M | M |
| 0.9 | C | C | C | C | M | M | M |
| 1 | C | C | C | C | M | M | M |

| p\d | 1 | 2 | 3 | 4 | 5 | 6 | 7 | 8 | 9 | 10 |
|---|---|---|---|---|---|---|---|---|---|---|
| 0.1 | C | C | C | C | C | C | C | C | C | C |
| 0.2 | C | C | C | C | C | C | C | C | C | M |
| 0.3 | C | C | C | C | C | C | C | C | M | M |
| 0.4 | C | C | C | C | C | C | C | M | M | M |
| 0.5 | C | C | C | C | C | C | M | M | M | M |
| 0.6 | C | C | C | C | C | M | M | M | M | M |
| 0.7 | C | C | C | C | C | M | M | M | M | M |
| 0.8 | C | C | C | C | C | M | M | M | M | M |
| 0.9 | C | C | C | C | M | M | M | M | M | M |
| 1 | C | C | C | C | M | M | M | M | M | M |

(a) $\tau = 0.1$   (b) $\tau = 0.5$

Fig. 14. Optimal policy construction.

section between $i$ and $j$ represents the action ($a$=C continuation, $a$=M migration of the service) to be taken by the FMC controller, where $i$ is the distance from the serving DC and $j$ is the probability $p$. We remark that the optimal policy construction has a threshold-based form, since beyond a certain distance value, the recommended action is service migration. This distance is inversely proportional to the probability $p$.

For instance, when $p = 0.8$, the proposed policy recommends service migration, if $d = 6$ and $d = 5$, for the first and second scenarios, respectively, since a high $p$ value means that the UE is moving far from the current DC, while low values indicate that the UE will most likely remain in the vicinity of the current DC, or will come back to the service area of the current DC. Furthermore, we remark that $\tau$ has an impact on the optimal policy construction, since there are less service migrations when $\tau$ is high. This is intuitive, as the incurred cost is not negligible compared to the achieved gain when migrating a service. In case of a random walk ($p = 0.5$), for both scenarios, the optimal policy recommends service migration when the distance exceeds 5, which represents a good tradeoff between cost and quality.

### 7.3 Testbed experiments

#### 7.3.1 OpenFlow-based implementation

We present early results obtained from experiments on our SDN-based FMC testbed described in Section 6.1.2. In particular, we focus on the evolution of service latency during and after the migration process, to demonstrate the advantages of FMC. To emulate increased service latency due to a user being served from a suboptimal DC, we introduced a 50 ms delay (round-trip) in the path between the client network and the suboptimal DC using Linux tc, while the RTT to the optimal DC was 1 ms. In our experiment, the client was originally being served from the suboptimal DC and we monitored the RTT from the user terminal to the VM hosting the service using ping. The red curve in Fig. 15 corresponds to a case when FMC is not used and the client keeps being served suboptimally, and thus the approximately 50 ms RTT. The green curve represents the case when FMC is active. When service migration is triggered, after a period of instability and a short-term increase in latency, the latter converges to a value close to 1 ms, after migration has been completed.

Note that the increased delay at the start of the experiment is mainly a result of the installation of OpenFlow rules when new traffic arrives at the FMCC. This implies a scalability issue for large centralized FMC deployments. Therefore, decentralization schemes, such as the one proposed by Bifulco et al. [15] could be considered to this end.

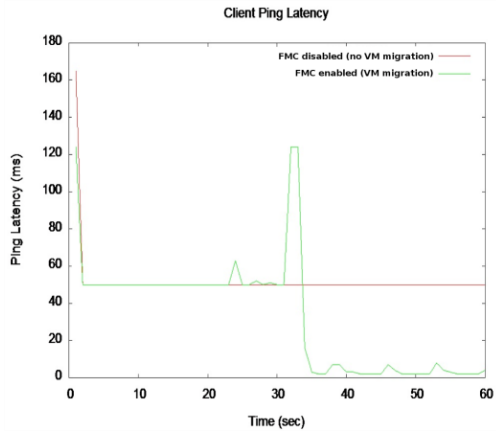

Fig. 15. Client ping latency with and without FMC.

### 7.3.2 LISP-based implementation

For our LISP-based FMC design, we experimented with the testbed described in Section 6.2.3. Our metrics of interest are (i) the service downtime duration, i.e., the time when the VM is not available, (ii) the RTT between the mobile user and the remote VM, and (iii) the migration duration.

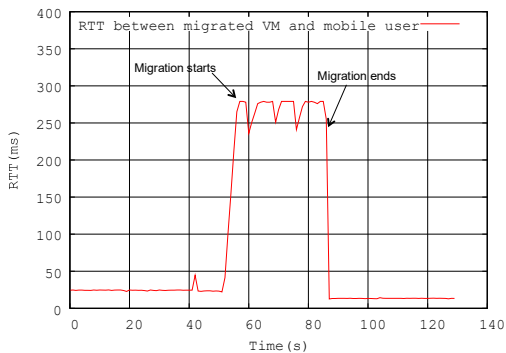

Fig. 16. The RTT between the migrated VM and mobile user.

Fig. 16 shows the instantaneous RTT between the migrated VM and the mobile user, measured using ping. Using Linux tc, we introduce a 10 ms RTT between the DCs. The RTT between the FMCC and xTR1 is approximately 1 ms, while the RTT between the FMCC and xTR2 is set to 10ms. At the start of the experiment, the user is connected to DC1, and enjoys good service quality ($RTT \approx 12$ ms). From $t = 55$ s, the user moves to another network and the measured RTT increases to approximately 250 ms. The FMCC thus triggers service migration to DC2, which is considered the optimal one. Migration starts at $t = 58$ s and ends at $t = 84$ s. During this period, the overall service downtime is 7.5 ms (not visible in the figure due to its short duration). This downtime is mainly due to the fact that the FMCC does not notify the involved xTRs about the change in the VM's EID until migration is completed. The kvm hypervisor at DC1 keeps the VM active until migration is complete and the VM keeps sending back ICMP echo replies from its original location (DC1). After migration is complete, the user is served by DC2 (optimal), which brings RTT down to a much lower value.

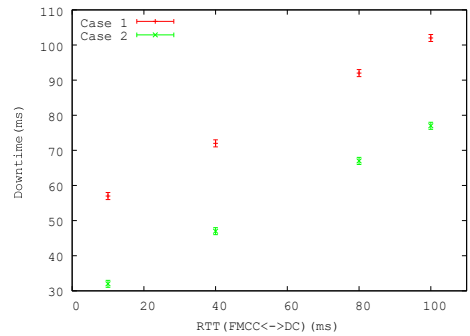

Fig. 17. Downtime duration as a function of path latencies. Each point is the mean value of a few tens of iterations.

We further quantify the dependence of the service downtime duration on the latencies in the paths between the entities of our architecture. In Fig. 17, we show the downtime duration when the VM is migrated from DC1 to DC2 for different values of the RTT between the FMCC and DC2 for the following testbed configurations: (i) The RTTs between the FMCC and xTR1 and xTR2 are set to 100 ms and 10 ms, respectively (case 1). (ii) The RTTs between the FMC controller and xTR1 and xTR2 are set to 50 ms and 50 ms, respectively (case 2).

It appears that the downtime duration is proportional to the RTT between the FMCC and the target DC, since the longer it takes to have the information from the DC about the success of a VM migration, the longer it takes to accordingly inform the xTR routers and thus redirect traffic to xTR2. We draw the same conclusion for the impact of the RTT between the FMCC and the xTR routers on downtime. Service downtime is mainly rooted at the LISP mobility management process, and its capacity to rapidly inform the xTR routers about the migration event. The VM size does not have a strong impact on downtime, since kvm activates the VM in DC2 only after migration is complete, which confirms the observations made by Raad et al. [23]. Importantly, the maximum downtime experienced is 100 ms (case 2), which is not expected to have a noticeable impact on service quality, especially for non-interactive applications.

Another conclusion from our experiments is that the duration of the service migration itself becomes practically independent of the RTT between DCs, as the latter increases. The time it takes to migrate a VM from DC1 to DC2 does not change much as the RTT in the DC1DC2 path increases beyond 10 ms. This is because VM migration is based on TCP, which is impacted more by the link bandwidth (set to 100 Mbps in this experiment) than by the RTT.

## 8 CONCLUSION

We presented our vision towards the enhanced delivery of cloud-based services to mobile users, tackling mobility-related challenges and offering an optimized user experience. We have designed Follow-Me Cloud, a framework that enables cloud services to "follow" users on the move, by performing sophisticated decisions to migrate service resources to the appropriate cloud infrastructure locations. We have defined an analytic model for the behavior of our system, and build on it to propose a Markov-Decision-Process-based algorithm for service migration decisions, which captures the tradeoff between migration cost and user experience. We have further developed two alternative FMC architecture designs, one which is based on SDN technologies and one making use of the LISP protocol. The presented numerical results from our model, as well as experiments with our SDN and LISP-based testbeds, demonstrate the potential of our approach for optimized mobile cloudbased service delivery and its feasibility for real-world deployment.

## APPENDIX LIST OF ACRONYMS

| | |
|---|---|
| CTMC | Continuous-Time Markov Chain |
| CTMDP | Continuous-Time Markov Decision Process |
| DTMDP | Discrete-Time Markov Decision Process |
| EID | Endpoint Identifier |
| ETR | Egress xTR |
| FMC | Follow-Me Cloud |
| FMCC | FMC Controller |
| ITR | Ingress xTR |
| MDP | Markov Decision Process |
| MR | Map Resolver |
| MS | Map Server |
| OF AR | OpenFlow Access Router |
| PGW | Packet Data Network Gateway |
| RLOC | Routing Locator |
| SGW | Serving Gateway |
| UE | User Equipment |
| xTR | Tunneling Router |

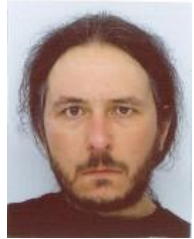

**Pantelis A. Frangoudis** received his B.Sc. (2003), M.Sc. (2005), and Ph.D. (2012) degrees in Computer Science from the Department of Informatics, AUEB, Greece. Currently, he is a post-doctoral researcher at team Dionysos, IRISA/INRIA Rennes, France, which he joined under an ERCIM post-doctoral fellowship (20122013). His research interests include wireless networking, Internet multimedia, network security, future Internet architectures, and QoE monitoring and management.

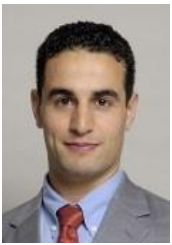

**Tarik Taleb** is currently a Professor at the School of Electrical Engineering, Aalto University, Finland. Previously, he was a senior researcher and 3GPP standards expert with NEC Europe Ltd., Heidelberg, Germany (2009-2015). Until March 2009, he was an assistant professor with the Graduate School of Information Sciences, Tohoku University. From October 2005 to March 2006, he was a research fellow with the Intelligent Cosmos Research Institute, Sendai. He received his B.E. degree (with distinction) in information engineering, and M.Sc. and Ph.D. degrees in information sciences from GSIS, Tohoku University, Sendai, Japan, in 2001, 2003, and 2005, respectively. His research interests include architectural enhancements to mobile core networks, cloud-based mobile networking, mobile multimedia streaming, inter-vehicular communications, and social media networking. Prof. Taleb has been also directly engaged in the development and standardization of the Evolved Packet System as a member of 3GPP's System Architecture working group.

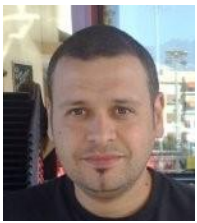

**Adlen Ksentini** is currently an Associate Professor with the University of Rennes 1, France. He is a member of the INRIA Rennes team Dionysos. He received the M.Sc. degree in telecommunication and multimedia networking from the University of Versailles, France, and the Ph.D. degree in computer science from the University of Cergy-Pontoise, France, in 2005, with a dissertation on QoS provisioning in IEEE 802.11-based networks. His other interests include future Internet networks, mobile networks, QoS, QoE, performance evaluation, and multimedia transmission.